\documentclass{article}\usepackage{natbib,emulateapj}
\usepackage{psfig}
\newcommand{\beq}{\begin{equation}}
\newcommand{\eeq}{\end{equation}}
\newcommand{\bea}{\begin{eqnarray}}
\newcommand{\eea}{\end{eqnarray}}
\newcommand{\bean}{\begin{eqnarray*}}
\newcommand{\eean}{\end{eqnarray*}}
\newcommand{\ba}{\begin{array}}
\newcommand{\ea}{\end{array}}
\newcommand{\bml}{\begin{mathletters}}
\newcommand{\eml}{\end{mathletters}}

\bibpunct[,]{(}{)}{;}{a}{}{,}
\begin{document}
\title{Hot Settling Accretion Flow onto a Spinning Black Hole}
\author{Mikhail V. Medvedev\altaffilmark{1} and Norman Murray}
\affil{Canadian Institute for Theoretical Astrophysics, 
University of Toronto, Toronto, ON, M5S 3H8, Canada }
\altaffiltext{1}{medvedev@cita.utoronto.ca; 
http://www.cita.utoronto.ca/$\sim$medvedev/}

\begin{abstract}
We study the structure and properties of hot MHD accretion onto 
a Kerr black hole. In such a system, the hole is magnetically coupled
to the inflowing gas and exerts a torque onto the accretion flow.
A hot settling flow can form around the hole and transport
the angular momentum outward, to the outer edge of the flow.
Unlike other hot flows, such as advection- and convection-dominated
flows and inflow-outflow solutions (ADAFs, CDAFs, and ADIOS), 
the properties of the hot settling
flow are determined by the spin of the central black hole, but
are insensitive to the mass accretion rate. Therefore, it may be possible 
to identify rapidly spinning BHs simply from their broad-band spectra.

Observationally, the hot settling flow around a Kerr hole is somewhat similar 
to other hot flows in that they all have hard, power-law spectra 
and relatively low luminosities. Thus, most black hole candidates in the 
low/hard and, perhaps, intermediate X-ray state may potentially accrete via 
the hot settling flow. However, a settling flow will be somewhat more 
luminous than ADAFs/CDAFs/ADIOS, will 
exhibit high variability in X-rays, and may have relativistic jets. 
This suggests that galactic microquasars and active galactic nuclei may 
be powered by hot settling flows. We identify several galactic X-ray
sources as the best candidates.
\end{abstract}

\section{Introduction}

Black holes (BHs) are identified both spectroscopically, by the
spectra of radiation emitted from an ambient accreting gas, and
via direct dynamical determination of their mass functions.
So far no measurement of the BH specific angular momentum, $j$, which 
describes the shape of the surrounding space-time, has been possible. 
The general relativistic effects associated with the Kerr metrics of a 
spinning BH are significant only within few gravitational radii from the
horizon; when observed from a large distance, Kerr and 
Schwarzchild space times look identical. This fact suggests 
that the BH spin can only be determined by studying 
the dynamics of the gas in the vicinity of the horizon.

Most spectroscopic BH candidates radiate significant levels of hard 
X-rays, indicating the presence of hot, optically-thin gas, which 
forms an advection-dominated, a convection-dominated accretion flow,
or an advection-dominated inflow-outflow solution (ADAF, CDAF, or ADIOS) 
\citep{NY94,NY95,NIA00,QG00,BB99}. The properties of such 
global, purely hydrodynamical flows are completely independent of $j$,
because the flow and a BH are dynamically decoupled: viscous stresses
vanish below the last stable orbit. In contrast, MHD flows can communicate 
via magnetic stresses as well, so that the situation is more subtle.

In this paper we demonstrate that the global structure and observational 
properties of an MHD accretion flow around a rapidly spinning BH are 
{\em determined} by $j$ in some cases. Therefore it may be possible to 
identify rapidly spinning BHs simply from their broad-band spectra.

Accretion flows are likely to contain rather strong magnetic fields 
produced via instabilities and/or MHD turbulence. At any instant as 
the gas falls into a BH, some magnetic field lines anchored at the inner 
part of the flow thread the horizon. A rapidly rotating BH in such an external 
magnetic field looses angular momentum and energy via a Blandford-Znajek-type 
(BZ) process \citep{BZ77,MT82}. The energy and angular momentum fluxes are 
transported along the field lines and damped in the inner part of an
accretion flow. For the accretion to proceed, this extra angular momentum
must be transported further away, to the outer boundary of the flow.
Neither an ADAF nor a CDAF are suitable, therefore.
One type of flow which supports such outward flux of angular momentum was
discovered recently by \citet{MN01}, as a model of accretion onto a 
spinning neutron star. The flow is referred to as the ``hot settling flow''.
(A second type of flow which could extract angular momentum is a jet or
ADIOS; see \citealp{BB99}.)

The rest of the paper is organized as follows. We begin by describing 
some relevant properties of the hot settling flow in \S \ref{S:VHSF}.
Formal derivation of the magnetically torqued flow is presented in 
\S\ref{S:BZSAF}. In \S\ref{S:OBS} we present observational predictions 
of the model and identify the objects for which the model may be applicable. 
\S\ref{S:DISC} presents discussion and conclusions.

\section{Viscous Hot Settling Flow}
\label{S:VHSF}

We briefly recall some basic properties of a hydrodynamical hot
settling flow onto a neutron star following the original work by 
\citet{MN01}; the effects of thermal conduction are discussed by \citet{MN01b}.

The hot settling flow solution  describes an optically
thin two-temperature accretion flow onto a rotating neutron star.  The
flow extracts the rotational energy of the star and radiates it via
free-free emission. The 
flow is geometrically thick with the vertical thickness being
comparable to the local radius of the flow, $H\sim R$. The flow is hot,
with the protons being at nearly the virial temperature. The flow is 
optically thin; energetically, it is cooling-dominated. The flow
rotates with a sub-Keplerian angular velocity.  The settling flow is 
powered by the rotational energy of the central star; hence, except for 
the radial velocity, {\em none} of the parameters of the flow
depend of the mass accretion rate $\dot M$. As in the case of an ADAF, 
this accretion rate must not exceed a critical value which is of order 
a few percent of the Eddington value; otherwise, the cooling is too strong 
and a thin Shakura-Sunyaev disk forms.

The settling flow consists of two zones: an inner two-temperature zone
for $r\la10^{2.5}$ ($r=R/R_S$, where $R_S=2GM/c^2$
is the Schwarzchild radius), and an outer one-temperature zone for
$r\ga 10^{2.5}$. Each zone is described by a self-similar solution. The
two- and one-temperature solutions read: 
\bea
& \rho=\hat\rho\, r^{-2},\quad
  \theta_p=\hat\theta_p\, r^{-1},\quad
  \theta_e=\hat\theta_e\, r^{p}, & ~ \nonumber\\
& \Omega=\hat\Omega\, r^{-3/2},\quad 
  v=\hat v\, r^0, & 
\label{ss-scaling}
\eea
where  $\theta_p=k_BT_p/m_pc^2$ and similarly for the electrons,
$\hat\rho,\ \hat\theta_p,\ \hat\theta_e,\ \hat\Omega,\ \hat v$ are
the density, temperatures, angular and radial velocities at the
BH horizon,
and the exponent $p=-1/2$ in the two-temperature zone and $p=-1$
in the one-temperature zone. Note that the ratio of the radial infall 
velocity to the free-fall velocity, $v_{\rm ff}=(GM/R)^{1/2}$, decreases 
with $R$ as $v/v_{\rm ff}\propto R^{1/2}$, hence the term ``settling''.
It is convenient here to normalize
$\Omega$ to the local Keplerian angular velocity 
$\Omega_K=(GM/R^3)^{1/2}$, that is 
\beq
s=\Omega/\Omega_K\propto\textrm{ const.}
\label{s}
\eeq
is the spin parameter. It is determined by the inner boundary 
condition, e.g., in the neutron star case, $s$ should match the 
spin of the star. In equation (\ref{ss-scaling}),
\bea
& \hat\rho\propto m^{-1}\alpha s^2, \quad
  \hat\theta_p\propto\textrm{ const.}, \quad
  \hat\theta_e\propto\textrm{ const.}, & \nonumber\\
& \hat\Omega\propto m^{-1}s, \quad
  \hat v\propto \dot m\alpha^{-1}s^{-2}, &
\eea
where $m=M/M_{\sun}$ is the mass of a central object in units of 
the solar mass, $\dot m=\dot M/\dot M_{\rm Edd}$ is the mass
accretion rate in Eddington units, 
$\dot M_{\rm Edd}=1.4\times10^{18}m\textrm{ g s}^{-1}$, and
$\alpha$ is the Shakura-Sunyaev viscosity parameter. As noted above,
none of the quantities (except for $v$) depend on the accretion rate
$\dot m$. The one-temperature self-similar solution is valid from
$r\sim10^{2.5}$ to approximately 
\beq
r_{ss}\sim\left(74\alpha^2s^2/\dot m\right)^2 
\sim 6\times10^3\alpha_{-1}^4 s_{-1}^4 \dot m_{-4}^{-2},
\label{r-ss}
\eeq
where $\alpha_{-1}=\alpha/10^{-1}$ and similarly for other quantities.
(Numerical solutions show that $r_{ss}$ is generally larger than 
the above estimate by a factor of ten or slightly less.)
Beyond this radius the solution is not self-similar.

The hot settling flow extracts rotational energy and angular
momentum from the central object. The angular momentum flux
is outward; it is constant through the entire accretion flow
and is equal to
\beq
\dot J_{\rm flow}=-6\pi c^2R_S^3\hat\rho\hat\theta_p\,\alpha\, s
\propto\alpha^2 s^3.
\label{J-dot-flow}
\eeq
The luminosity of the hot settling flow is due to both the
gravitational energy of the infalling gas (the term 
$\propto\dot m$) and the rotational energy that is released as 
the central star is braked by viscosity 
(the term independent of $\dot m$):
\bea
L_{\rm flow}&\simeq&3.2\times10^{34}mr^{-1}\dot m_{-2}s^2_{-1} \nonumber\\
& &{ }+2.7\times10^{34}mr^{-1}\alpha_{-1}^2s_{-1}^4 \textrm{ erg s}^{-1}.
\label{L}
\eea
Note that the rotational luminosity is a strong function of $s$.
The radiative efficiency of the flow, $\eta=L_{\rm flow}/\dot Mc^2$
may be arbitrarily large for low $\dot M$ -- a result of the fact that 
there are two energy sources. Also, the radiative 
efficiency of a hot settling flow for a given $\dot M$ is larger than
the efficiency of a hot flow onto a non-spinning object. 
Another feature of the
settling flow (as a result of large density) is that optically thin 
Bremsstrahlung cooling (which is sensitive to $\rho$) dominates over
self-absorbed synchrotron cooling; this is the result of larger 
$\rho$ in the flow. Thus, the spectrum $L_\nu$ 
is flat, with a high energy cutoff. At a typical X-ray energy, 
$h\nu\sim 10$~keV, the observed luminosity per decade is
\beq
\nu L_\nu\simeq5.6\times10^{-3}L_{\rm flow}(h\nu/10~{\rm keV}).
\label{nuLnu}
\eeq
The spectrum has a cutoff in the range of $\sim$~few hundred keV to few MeV.
The above estimate of the shape of the spectrum is rather crude. It 
does not include synchrotron emission and Comptonization, the latter 
should be very strong in the very hot inner part of the flow. We do not 
attempt to calculate accurate spectral models of the hot settling flow. 

For reasonable values of the adiabatic index $\gamma\ge1.5$,
the Bernoulli number of the flow is negative; therefore, in the 
absence of the dynamically important magnetic fields, a strong 
outflow or a wind is not expected. The hot settling flow is also 
convectively stable if $\gamma\ge1.5$ \citep{MN01}. 
The flow is cooling-dominated and, hence, might be thermally 
unstable. However, \citet{MN01b} demonstrated that thermal conduction 
in the flow is strong enough to make the flow marginally stable.

Numerical simulations (see \citealp{MN01}) indicate that the above 
scalings are accurate for the values of spin $s\la0.5$. However, as 
the spin is decreased below $s\sim0.1$, the hot settling solution 
smoothly transforms to an ADAF-type 
solution, which becomes well-established for $s\la0.01$. The transition 
is not sharp, so that the value of $s_t$ at which the transition occurs
is difficult to identify. Numerical experiments indicate that the value 
of $s_t$ is not sensitive to the outer and inner radii of the flow,
$\gamma$, and $\dot M$ and is, roughly, $s_t\sim0.04-0.06$. Below we
relate the spin parameter $s$ to the dimensionless angular velocity
of a black hole, $j$.

\section{BZ Powered Settling Accretion Flow}
\label{S:BZSAF}
\subsection{Preliminary considerations}

The crucial difference between accretion onto a neutron star and 
onto a black hole is the existence of the hard surface in the former case.
As the gas falls onto a spinning star, it piles up and settles onto its 
surface, forming a narrow boundary layer. The viscous torque in this layer 
brakes the star and imposes a no-slip boundary condition for the accretion
flow. In contrast, in a BH case, the viscous torque vanishes identically 
at the horizon and the viscous stress must be very small below the last 
stable orbit (see e.g., \citealp{B70,T74,AK89}). Therefore, a spinning BH 
cannot communicate to an accretion flow by viscosity alone. This imposes 
a no-torque boundary condition. Hence, a purely viscous hot settling flow 
onto a BH is not possible.

Magnetic fields, which are commonly present in accretion flows, change
the situation drastically. Turbulent magnetic fields create an average 
Maxwell stress which does not vanish at the last stable orbit \citep{G99,K01}.
More interestingly, these turbulent magnetic loops, advected with the 
material and continuously crossing the horizon, provide magnetic coupling 
between the flow and a BH, as sketched in Figure \ref{fig1}. 
The energy and angular momentum are extracted
from a BH via a Blandford-Znajek-like  mechanism \citep{BZ77,MT82} and 
transported along field lines to the accretion flow in the form of the 
Poynting flux. The continuity of the angular momentum flux replaces the 
no-torque inner boundary condition for the flow. As in the neutron star case 
we will parameterize this boundary condition with an {\em effective} spin 
of the central object, $s_{\rm eff}$.

\subsection{Quantitative analysis}

A spinning (Kerr) black hole interacting with an external
magnetic field behaves as an imperfect rotating conductor.
It generates currents flowing along magnetic field lines
which constitute a force-free magnetosphere around the BH.
The energy and angular momentum fluxes are directed away 
from the BH horizon, what indicates that the BH is 
spun down. The geometry of the magnetic field far from 
the horizon determines where this energy and angular
momentum are ultimately deposited. Open field lines allow
them to, practically, escape to infinity. The energy then 
powers a jet outflow and the angular momentum is carried away 
by the outflowing gas. In contrast, if magnetic field lines 
are anchored at the inner accretion flow, the energy and 
angular momentum extracted from the hole are deposited into 
the accreting gas. This latter case is of interest to us.

We now briefly sketch the derivation of the BZ angular momentum 
losses. We consider a Kerr BH in the Boyer-Lindquist coordinates
$x^\mu=(t,r,\theta,\phi)$,
\bea
ds^2&=&\left(1-\frac{2Mr}{\Sigma}\right)\,dt^2
+\frac{4Mar}{\Sigma}\,\sin^2\theta\,dtd\phi  \nonumber\\
& &{ }-\frac{\Sigma}{\Delta}\,dr^2-\Sigma\,d\theta^2
-\frac{\Xi}{\Sigma}\,\sin^2\theta\,d\phi^2,
\eea
where 
\bean
\Sigma&=&r^2+a^2\cos^2\theta,\\ 
\Delta&=&r^2-2Mr+a^2=(r-r_+)(r-r_-), \\
\Xi&=&\left(r^2+a^2\right)^2-\Delta\,a^2\,\sin^2\theta, 
\eean
$a=J_H/M$ is the BH angular momentum per unit mass, 
$r_+=M+\sqrt{M^2-a^2}$ is the radius of the event horizon,
and we use natural units, $c=G=1$. We assume steady state 
and axisymmetry, hence $\partial_t=\partial_\phi=0$. 
As in any axially symmetric system, 
the flux vector for angular momentum about the axis of symmetry 
is conserved, i.e., there is a corresponding axial Killing 
vector $\xi^\mu=(0,0,0,1)$. The total energy-momentum tensor 
of the electromagnetic field is defined as
\beq
T^{\mu\nu}=\frac{1}{16\pi}\,g^{\mu\nu}F_{\alpha\beta}F^{\alpha\beta}
-\frac{1}{4\pi}\,g^{\mu\alpha}F_{\alpha\beta}F^{\nu\beta},
\eeq
where $F_{\mu\nu}=A_{\nu,\mu}-A_{\mu,\nu}$ is the electromagnetic 
tensor, $A_\mu$ is a 4-vector potential, and the comma denotes the 
partial derivative with respect to $x^\mu$. From the conservation law
\beq
T^{\mu\nu}_{~~;\nu}=0
\eeq
and the Killing equation $\xi_{\mu;\nu}+\xi_{\nu;\mu}=0$, 
one has the following conservation law 
\beq
(T^{\mu\nu}\xi_\mu)_{;\nu}=0
\eeq
(the semicolon denotes the covariant derivative). This equation represents 
a conservation of flux $T^{\mu\nu}\xi_\mu$. Substituting $\xi^\mu$ from 
above, we see that the conserved angular momentum flux is the 
$\phi$-component of the energy-momentum tensor
\beq
{\cal L}^\mu=T^{\mu\nu}\xi^\alpha g_{\alpha\nu}
=T^\mu_{~\alpha}\xi^\alpha=T^\mu_{~\phi}.
\eeq
The outgoing flux is given by the $r$-component of ${\cal L}^\mu$ as
\bea
\dot{\cal J}\equiv T^r_{~\phi}
&=&-\frac{1}{4\pi}\left(F^r_{~\alpha}F_\phi^{~\alpha}\right)
=-\frac{1}{4\pi}\,A_{\phi,\theta}B_\phi\, g^{rr}g^{\theta\theta}
\nonumber\\
&=&-\frac{1}{4\pi}\,A_{\phi,\theta}\,B_\phi\frac{\Delta}{\Sigma^2},
\label{BZ}
\eea
where $B_\phi=F_{r\theta}=A_{\theta,r}-A_{r,\theta}$, is the toroidal 
covariant component of the magnetic field which arises because of frame 
dragging near the BH horizon. At the horizon, the stream function $A_\phi$ 
must be finite \citep{Z77,MT82} and, hence,
\beq
B_\phi=\Delta^{-1}\left[\Omega_F\left(r_+^2+a^2\right)-a\right]
\,A_{\phi,\theta},
\label{condition}
\eeq
where $\Omega_F(r,\theta)=-A_{0,r}/A_{\phi,r}=-A_{0,\theta}/A_{\phi,\theta}$
measures the angular velocity of magnetic field lines which is also due to 
the effect of relativistic dragging of inertial frames. Equation 
(\ref{condition}) relates at the horizon the components of the magnetic
field which are normal, $B_r\propto A_{\phi,\theta}$, and tangential, $B_\phi$, 
to the BH horizon. Equations (\ref{BZ}), (\ref{condition}) now yield
\beq
\dot{\cal J}
=A_{\phi,\theta}^2\,
\frac{\left(r_+^2+a^2\right)\left(\Omega_H-\Omega_F\right)}
{4\pi\,\left(r_+^2+a^2\cos^2\theta\right)^2},
\eeq
where $\Omega_H\equiv a/(r_+^2+a^2)$ is the BH angular velocity
and $\left(r_+^2+a^2\right)=2Mr_+$.
This equation represents the angular momentum flux coming from a BH per 
the unit area of the horizon. The total angular momentum extraction
rate is obtained integrating $\dot{\cal J}$ over the horizon area,
and scales approximately as
\beq
\dot J_{\rm BZ}=
\int_{\cal H}\dot{\cal J}\,\Sigma\sin\theta\,d\theta\, d\phi\,
\simeq C\,B_\bot^2 r_+^4\left(\Omega_H-\Omega_F\right),
\label{J-tot}
\eeq
where we use  $B_\bot(\theta)=\left[\Xi^{1/2}(r_+)\sin\theta
\right]^{-1}A_{\phi,\theta} =\left[(r_+^2+a^2)\sin\theta\right]^{-1}
A_{\phi,\theta}$, we denoted $B_\bot$ as an average strength of the 
magnetic field normal to the BH horizon \citep{Lee+01,TPMc86},
 and $C$ is a constant which value depends on the magnetic field 
geometry in the force-free region. \citet{MS96} estimated this
constant (see also \citealp{MT82}) to be $C\simeq1/8$.

The rate of the angular momentum extraction in the BZ process is given 
by equation (\ref{J-tot}). This equation may be written in a more
convenient form. First of all, we restore dimensional $c$ and $G$.
Next, by analogy with equation (\ref{s}) we define the dimensionless 
BH spin and angular velocity of the field as
\beq
\omega_H=\Omega_H/\Omega_{\rm max}, \qquad \omega_F=\Omega_F/\Omega_{\rm max},
\eeq
where $\Omega_{\rm max}=c^3/GM$ and $\Omega_H=ac^3/2GMr_+$. 
The dimensionless horizon radius of a BH is 
\beq
r_H=\frac{r_+}{R_S}=\frac{r_+c^2}{2GM}
=\frac{1}{2}\left[1+(1-j^2)^{1/2}\right],
\eeq
where $j=J_H/J_{\rm max}=J_Hc/GM^2$ is the dimensionless angular
momentum of a BH (i.e., $j=a/M$ in $c=G=1$ units); note that 
$j=4r_H\omega_H$. We also introduce $C_{1/8}=C/(1/8)$ for convenience.
With all these definitions, the rate of loss of angular momentum 
by a BH, which is $\dot J_{\rm BH}=-\dot J_{\rm BZ}$, becomes
\beq
\dot J_{\rm BH}\simeq-(1/4)C_{1/8}B_\bot^2R_S^3r_H^4(\omega_H-\omega_F).
\label{J-dot0}
\eeq

The magnetic field strength near the BH horizon is not generally known. 
However, if the field is produced in an accretion flow, e.g., via the 
magneto-rotational instability, it will likely be nearly in equipartition 
with the gas energy. Hence, we parameterize it as follows
\beq
\frac{B_\bot^2}{8\pi\rho c_s^2}
=\frac{B_\bot^2}{8\pi c^2\hat\rho\hat\theta_p r_{\rm in}^{-3}}\equiv b^2\la1,
\eeq
where we have used the self-similar solution for the hot settling flow
(\ref{ss-scaling}),  $\hat\rho$ and $\hat\theta_p$ are constants and
the inner boundary of the accretion flow is at $r_{\rm in}$. Then
equation (\ref{J-dot0}) becomes
\beq
\dot J_{\rm BH}\simeq
-\frac{\pi}{2}C_{1/8}b^2R_S^3c^2\hat\rho\hat\theta_p j
\left(\frac{r_H}{r_{\rm in}}\right)^3\left(1-\frac{\omega_F}{\omega_H}\right).
\label{J-dot}
\eeq

To account for a general geometry of the magnetic field in the system
BH~+~accretion flow, we assume that there are both open field lines
which extend to infinity and form a jet and that there are field lines which 
thread the inner boundary of an accretion flow
(a well-defined inner boundary is an idealization, of course).
Then only a fraction $\epsilon\le1$ of the extracted angular momentum goes
into the flow and $(1-\epsilon)$ is ``lost'' through the open field
lines and powers the jet. In steady-state, the angular momentum flux
from the hole to the flow, $\epsilon \dot J_{\rm BH}$, must be equal to 
the flux through the viscous flow itself given by equation (\ref{J-dot-flow}),
that is $\dot J_{\rm flow}=\epsilon \dot J_{\rm BH}$. This condition yields
\beq
\alpha s_{\rm eff}\simeq\frac{C_{1/8}}{12}\,\epsilon\,b^2\,j\, 
\left(\frac{r_H}{r_{\rm in}}\right)^3\left(1-\frac{\omega_F}{\omega_H}\right).
\label{main}
\eeq
The rotation velocity of field lines $\omega_F$, entering equation 
(\ref{main}), is determined by the conditions in the force-free region
and in the accretion flow. \citet{MT82} argued that the field in the
force-free region will adjust itself so as to maximize the power output,
that is $\omega_F\simeq\omega_H/2$. 

Finally, it is instructive to estimate $ s_{\rm eff}$ for a nearly 
extremal Kerr BH, $j\sim1$. The inner radius of the accretion flow, 
which is comparable to the radius of the last stable orbit, is 
$r_{\rm in}\sim r_{\rm lso}\sim r_H$ in that case. We assume ``violent'' 
MHD turbulence with equipartition between the 
kinetic and magnetic energy, $b\sim1$, and a $100\%$ efficiency
of the BZ torque, $\epsilon\sim1$. The Shakura-Sunyaev viscosity 
parameter $\alpha$ in an 
accretion flow typically ranges from $\sim0.1$ to 0.3; 
we assume $\alpha\sim0.1$. Then, the effective spin 
parameter of the hot settling flow around a maximally rotating BH is 
\beq
s_{\rm eff,max}\sim0.4\,.
\eeq
This value is large enough to be confident that the hot settling 
flow can exist in a BH system. However, because of the strong 
dependence of the magnetic BZ-like torque on $r_{\rm in}$ 
and $b$, one can expect that fluctuations of these quantities in the 
accretion flow can change the type of flow, i.e., switch between the 
settling flow and an ADAF/CDAF. This will result in strong variability
of the system.

\section{Observational predictions}
\label{S:OBS}

A hot settling flow and an ADAF/CDAF form a single sequence of solutions
in which the spin parameter, $s_{\rm eff}$, plays a pivotal role of 
a control parameter. These types of flow are quite similar: both
have very hot (nearly virial), optically-thin gas, forming a quasi-spherical
accretion flow with very low mass accretion rate. The observational
properties are also expected to not be drastically different: both flows 
have hard, power-law spectra extending to at least few hundred keV, 
cf. equation (\ref{nuLnu}), and relatively low (sub-Eddington) luminosities.
These radiation properties suggest that the BH candidates 
in the low/hard state, e.g., low-mass X-ray binaries (LMXBs) with 
radiatively inefficient (well below Eddington) accretion, may be 
good candidates for a hot settling flow type accretion too.

Despite similarities, a hot settling flow is not identical to other hot flows. 
First, the hot settling flow is more radiatively efficient and, hence, more 
luminous than an ADAF, because it is powered by rotation of the central object.
The maximum luminosity, estimated from (\ref{L}) for a nearly extremal BH
with $r_{\rm in}\sim r_H\sim1/2$, is
\bea
L_{\rm max}&\simeq&1.6\times10^{37}\,
\frac{C_{1/8}^4\,m}{\alpha_{-1}^{2}\,r_{{\rm in},1/2}}\,\epsilon^4\,j^4\,b^8\,
\left(\frac{r_H}{r_{\rm in}}\right)^{12}~\textrm{erg s}^{-1}
\nonumber\\
&\simeq&0.13\,L_{\rm Edd}\,
\frac{C_{1/8}^4\,\epsilon^4\,j^4\,b^8}{\alpha_{-1}^{2}\,r_{{\rm in},1/2}}\,
\left(\frac{r_H}{r_{\rm in}}\right)^{12}
\label{L-max}
\eea
where $b\simeq1$,\ $\epsilon\sim1$,\  $r_{{\rm in},1/2}=r_{\rm in}/(1/2)$
and $L_{\rm Edd}=1.25\times10^{38}m~\textrm{erg s}^{-1}$ 
is the Eddington luminosity. Second, $L$ is a very strong function of 
the effective spin, which, in turn, is sensitive to $b$ and $r_{\rm in}$.
Even a mild fluctuation of either of the two will result in a huge 
variation of the radiation flux. We should emphasize that the luminosity 
of the flow is practically independent of the mass accretion rate 
(especially for high $s_{\rm eff}$); $\dot M$ should not exceed few percent 
of the Eddington accretion rate in a settling flow. It is interesting that 
the variability of the source
is associated not with varying $\dot M$, but with the level of turbulence 
and the geometry of the flow close to a BH. Third, the hot settling flow 
can occur around rapidly rotating BHs only. Such BHs may be
responsible for the production of collimated relativistic outflows 
(jets), observed in some systems. Thus, higher luminosity and variability 
of an object and the existence of jets allow us to identify two 
classes of BH candidates which may accrete via the hot settling flow,
namely microquasars and active galactic nuclei.

\subsection{The Galactic microquasars}

The Galactic microquasars are relatively bright, often highly 
variable objects with resolved jets, some of which show superluminal 
motion of their sub-components (see the review by \citealp{MR99}).
Some of these systems exhibit quasi-periodic oscillations (QPOs), the 
high-frequency QPOs usually correlate with the hard power-law spectral 
component emerging in the very high and/or high state
(see, e.g., \citealp{Remillard+99,Homan+01}).

GRO~J1655-40 is the most interesting object in the context of settling 
flows. It is a highly variable LMXB 
with $M_{\rm BH}\sim7M_{\sun}$ \citep{Shahbaz+99}, the X-ray luminosity 
changes from $L_X\ga0.2L_{\rm Edd}$ with a strong power-law component in a 
very high state to $L_X\sim0.01L_{\rm Edd}$ in quiescence \citep{Remillard+99}. 
Several QPOs have been detected in the system. \citet{Cui+98} interpreted the 
high-frequency QPO at $\nu=300$~Hz as due to precession of an accretion 
disk caused by the inertial frame dragging; this suggests 
a nearly maximally rotating central BH. \citet{Zhang+97} obtained 
a similar result considering X-ray properties of the inner accretion disk.
However, \citet{Sobczak+99} considered uncertainties of the model
and put an alternative constraint on the BH angular momentum  $j<0.7$.
The recent discovery of a second high-frequency QPO at 450~Hz 
\citep{Strohmayer01} indeed implies a spinning BH. Assuming that the QPO
frequency is set by the orbital frequency at the last stable orbit,
one obtains $0.15<j<0.5$ for a range of BH masses $5.5-7.9\,M_{\sun}$,
which are the $95\%$ confidence limits from \citet{Shahbaz+99}. 
The existence of a central, rapidly spinning BH is, thus, almost certain.
Therefore, the object seems to be the best candidate for a hot settling
accretion flow. 

The following three objects are also good candidates for the hot settling 
flow accretion; all are highly variable X-ray binaries and may have rapidly 
spinning BHs as primaries.

GRS 1915+105 is similar to GRO~J1655-40. It is a recurrent X-ray transient, 
possibly LMXB \citep{Greiner+01b}. The mass of the BH is $14\pm4\,M_{\sun}$
\citep{Greiner+01a}. 
The highest-frequency QPO has $\nu$=67~Hz. If one interprets this QPO as
an X-ray modulation at the precession frequency of an accretion disk
because of relativistic frame dragging by a spinning BH, 
the angular momentum of the central BH would be $j\sim0.65$ and $j\sim0.95$
for a $3\,M_{\sun}$ and $30\,M_{\sun}$ BH, respectively  \citep{Cui+98}.
  
XTE~J1748-288 is another transient X-ray source. This is a binary with
a low-mass companion \citep{Liu+01} and the primary of mass $>4.5\,M_{\sun}$ 
\citep{Miller+01}. Several QPOs (the highest frequency is 30~Hz) were observed 
when this source was in the very high and high states and none were observed 
in the low state \citep{Revnivtsev+00}. An Fe K$\alpha$ fluorescence 
line observed in the very high state appears broad and strongly redshifted,
consistent with the emission from the last stable orbit around a 
maximally rotating BH \citep{Miller+01}.

XTE~J1550-564 is an X-ray transient with a BH primary of mass
$8.4\,M_{\sun}\la M_{BH}\la 11.6\,M_{\sun}$ \citep{Orosz+01}.
It is a highly variable source; the X-ray luminosity varies by a factor of
1000 \citep{Homan+01}. The detection of high-frequency QPOs with $\nu$
up to 284~Hz \citep{Homan+01,Miller+01b} may indicate a spinning central BH.

There are also several persistent X-ray sources with typical BH 
low/hard spectra and radio jets which are worthwhile mentioning, namely 
Cyg~X-1, GRS~1758-258, and 1E~1740.7-2942. Cygnus~X-1 (also known as
(HDE~226868 and V1357~Cygni) is a strong BH candidate with a primary
mass between $5-15M_{\sun}$ \citep{Herrero+95} and a supergiant secondary 
with mass between $20-33M_{\sun}$. For most of the time it is observed 
in the low/hard X-ray state. Recently a relativistic radio jet has been 
detected \citep{Stirling+01}. The sources GRS~1758-258 and 1E~1740.7-2942 
are located in the  bulge; their companions are unknown because of high 
Galactic extinction in the optical, though  high mass 
companions are unlikely \citep{Mereghetti+97,Marti+98,Marti+00}. 
Both systems have radio jets but do not reveal superluminal motion. 
Both sources are variable in X-rays with the maximum luminosity 
$L_{\rm max}\sim3\times10^{37}~\textrm{erg s}^{-1}$ \citep{Main+99}. 
The X-ray spectra of these sources are typical of BH systems in the low/hard 
state. At large energies, they are dominated by a hard power-law
component extending up to $\sim300$~keV \citep{Main+99,Lin+00}. 
Only low-frequency QPOs with $\nu\le1$~Hz have been detected in GRS~1758-258.
This, however, cannot exclude a spinning central BH with very ``laminar''
accretion and/or weak magnetic coupling between the BH and the 
accretion flow at the present time.

\subsection{Active galactic nuclei}

There is strong dynamical evidence that galactic nuclei contain 
supermassive BHs with masses in the range from few times $10^6$ 
to few times $10^9\,M_{\sun}$. From the analysis of the contribution 
of quasars to the X-ray background in the universe, \citet{Elvis+02}
demonstrate that the accretion efficiency must be greater than 
$\sim15\%$, which is well above the maximum efficiency of a flow
around a non-spinning BH. This result implies that most supermassive 
BHs might be spinning. Galactic collimated outflows, the best example 
being the optical jet of M87, may be indicative of a spinning BH too.

We speculate that some AGNs can be candidates
for the hot settling flows. In this case, the high observed luminosities
must be attributed to the strong BH--accretion flow coupling, rather
than the high mass accretion rate. Clearly, not all AGNs will fall
into this category. Moreover, even if the hot settling flow forms near 
a central BH, its radiation may be strongly contaminated by radiation 
from an outer accretion disk/torus. Whether AGNs contain rapidly rotating
BHs is not clear. However, the jet activity and temporal variability 
may indicate this. The most convincing case, presented to date for
having measured the BH spin has been given for the Seyfert~1 galaxy 
MCG~--6-30-15 \citep{Tanaka+95,BF01,Wilms+01} where the measured very 
broad and highly redshifted shape of the Fe K$\alpha$ fluorescent line 
profile suggests that it may be produced in an accretion disk at the 
last stable orbit around 
a nearly maximally spinning BH. However, this is not the only possible 
interpretation \citep{RB97}. Recently \citep{Pariev+01} presented 
a much more detailed analysis of iron line emission from an inner 
accretion disk in the Kerr geometry. They have found that the data 
for MCG~--6-30-15 are consistent with a thick (and, hence, hot) 
accretion disk, whereas the thin Shakura-Sunyaev disk is ruled out
at $3\,\sigma$, and that the central BH must spin with $j\ga0.26$.

Another class of candidates are central BHs in low-ionization nuclear
emission line region (LINER) galaxies \citep{A-H+99,Komossa+99}.
These galaxies often have a compact X-ray nucleus and their 
luminosities range from $\sim5\times10^{39}$ to $\sim5\times10^{41}$
erg~s$^{-1}$, which is $\sim1$ to 3 orders of magnitude smaller 
than typical Seyferts. These facts together indicate the presence of 
a central supermassive BH accreting at low (highly sub-Eddington)
rates via some sort of a hot accretion flow. A good candidate is,
for instance, NGC~4579 which has a power-law spectrum with the photon
index $\Gamma\sim1.2$, consistent with Bremsstrahlung emission
\citep{Eracleous+01} but steeper than what an ADAF model predicts
\citep{NYM96}.

\section{Discussion and conclusion}
\label{S:DISC}

The conventional folklore of accretion onto a BH dictates that 
the properties and structure of an accretion flow do not depend
on whether the BH is rotating or not. If so, then at low $\dot M$ only 
the ADAF, CDAF or ADIOS are the relevant solutions of the BH accretion problem.
In this paper we demonstrated that the folklore is, in general,
wrong. The magnetic coupling between a spinning BH and a low $\dot M$
hot  accretion flow can
cause a different type of flow to form, namely a hot settling flow.
This flow can be distinguished from other hot flows by its 
kinematic and broad-band spectral properties, which depend on 
the spin of the central object and the strength of the magnetic 
coupling, but which are quite insensitive to the mass accretion rate. 
We should note here an interesting model of a thin disk torqued by 
a BH \citep{LP00,L02}. Unlike our model, however, this coupling affects 
the flow near the last stable orbit only, whereas the global 
structure of the Shakura-Sunyaev disk remains unchanged. More recently,
Li has generalized the model to allow the magnetic field to couple to 
the disk over a range of radii. However, these models all assume that 
the flow is radiatively efficient, in contrast to the hot flows
considered here.

There are two issues which we did not discuss in the paper. First,
one can imagine that strong shearing motion inside the last stable 
orbit could result in tight winding of magnetic field lines and
their further reconnection \citep{B00}. Indeed, in the thin disk case
considered by Li, the magnetic field is localized in the plane of the 
disk, so that reconnection can easily take place. In contrast, the hot 
settling flow is geometrically thick; thus different field lines are threading
the horizon at various latitudes. This geometry makes the field more difficult
to reconnect. Second, the flow is continuously advecting the field.
The infall velocity in hot flows is usually larger than the
radial velocity of the gas in the Shakura-Sunyaev disk (in the
settling flow $v$ is set by $\dot M$ which, therefore, should not
be too small). Because of shorter infall time, magnetic 
reconnection in hot flows should be less likely.
Third, along with angular momentum flux from a hole, there is 
an energy flux, $\dot{\cal E}=\Omega_F\dot{\cal J}$. This extra energy 
is deposited at the inner edge of the flow. The structure of this very 
hot inner ``boundary layer'' may be quite complicated because it
is likely affected by Compton cooling, thermal conduction, etc..
This problem is beyond the scope of our paper.

The hot settling flow around a Kerr hole is observationally similar 
to other hot flows (ADAFs, CDAFs, and ADIOS) in that all they have hard,
power-law spectra, extending to, at least, a few hundred keV and 
relatively low (sub-Eddington) luminosities. This suggests that 
most BH candidates in the low/hard X-ray state may potentially
accrete via the hot settling flow. 
However, the sources accreting via the settling flow must
be (i) more luminous on average (as there is an extra source of energy),
(ii) highly variable in X-rays, and (iii) may have relativistic jets.
(There is a possible caveat here that jets may be launched by the 
inner accretion disk and not by a spinning BH.) 
Based on these predictions  we have identified two classes of 
objects which could be good candidates for hot settling flows, namely 
galactic microquasars and low luminosity active galactic nuclei.

\acknowledgements 

The authors are grateful to Ramesh Narayan, Roger Blandford, 
and Peter Goldreich for discussions and valuable comments.
This work was supported by the CITA fellowship.

\figcaption[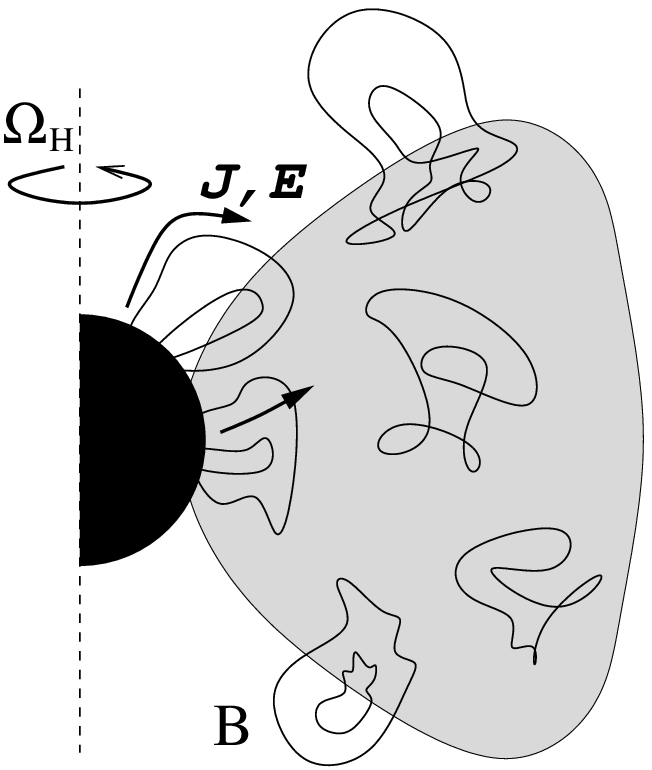]{Cartoon showing magnetic couping of a black hole 
to a turbulent magnetized flow.
\label{fig1} }

\bigskip
\psfig{file=BZ.eps}

\end{document}